\documentclass{desyproc}

\begin{document}

\title{Testing a new WISP Model with Laboratory Experiments}

\author{{\slshape Pedro Alvarez$^1$, Paola Arias$^2$, Carlos Maldonado$^2$\\[2ex]
$^1$Departamento de Fisica, Universidad de Antofagasta, Aptdo 02800, Chile.\\
$^2$Departamento de F\'isica, Universidad de Santiago de Chile, Santiago, Chile.}}

\contribID{Arias\_Paola}

\confID{16884}  
\desyproc{DESY-PROC-2017-02}
\acronym{Patras 2017} 
\doi  

\maketitle

\begin{abstract}
We explore the phenomenological consequences of a model with axion-like particles and hidden photons mixing with photons. In this model, the hidden photon is directly coupled to the photon, while the axion coupling is induced by an external electromagnetic field. We consider  vacuum effects on a polarised photon beam, like changes in the ellipticity and rotation angles.
\end{abstract}

\section{Introduction}

In this work we would like to go beyond the straightforward extension of the Standard Model, namely the  one-missing-particle paradigm. On the one hand, seems timely, due to the extraordinary refinement in sensitivity of latest years of experiments looking for WISPs (Weakly Interactive Slim Particles). On the other hand, there are some proposals of more complex models with rich phenomenology, such as \cite{Masso:2005ym}, where they consider a model with a hidden photon (HP) coupled to an axion-like field (ALP), and the kinetic mixing term. Also in \cite{Jaeckel:2014qea} a model with axion-like particle + hidden photon was invoked to explain the 3.55 keV line in the spectra of galaxy clusters. More recently a similar model was considered \cite{Kaneta:2016wvf}, where the pseudo-scalar boson is the QCD axion, which is coupled to the hidden photon.
We have chosen to follow the construct in \cite{Jaeckel:2014qea}, therefore the hidden photon is the mediator between visible and hidden sector. In this work we are interested in observables effects of this model, focusing on vacuum effects, like dichroism and birefringence.

\section{The model and equations of motion}

We consider the following effective Lagrangian:
\begin{equation}
\mathcal L=-\frac{1}{4}f_{\mu\nu}f^{\mu\nu} - \frac{1}{4}x_{\mu\nu}x^{\mu\nu} + \frac{1}{2}\partial_\mu\phi\partial^\mu\phi + \frac{1}{2}\sin\chi f_{\mu\nu}x^{\mu\nu} 
+ \frac{1}{4}g\phi x_{\mu\nu}\tilde{x}^{\mu\nu}  
- \frac{m_\phi^2}{2}\phi^2  
+ \frac{m^2_{\gamma'}\cos^2\chi }{2}x_\mu x^\mu.
\notag
\end{equation}
Here   $a_\mu$ is the photon field, $(x_{\mu\nu})$ the HP field and $\phi$ is the ALP. The HP is directly coupled to photons via the kinetic mixing term, parametrised by $\sin\chi$. As it is well known, defining $X_{\mu}= x_{\mu}- a_\mu \sin \chi$ and $A_{\mu}= a_{\mu} \cos\chi$, removes the kinetic mixing, but at the price to inherit a coupling in the mass sector between photons and HPs, and also a term of the form $g\tan^2\chi\phi F_{\mu\nu}\tilde{F}^{\mu\nu}$, meaning an explicit coupling between photons and ALPs.

We start assuming  a photon beam source of frequency $\omega$, propagating in $z$ direction, and the plane wave approximation, i.e, $\vec{A}(z,t)=e^{i\omega t}\vec{A}(z)$, $\phi (z,t)=e^{i\omega t}\phi (z)$ and $\vec{X}(z,t)=e^{i\omega t}\vec{X}(z)$. We also include an homogeneous magnetic field $\vec B$, oriented in the $\hat x$ direction. Additionally, we assume no external hidden fields are present. Secondly, we linearise the equations of motion assuming the external electromagnetic field is much stronger than the photon source $|\vec A_{\rm ext}| \gg |\vec A| $, and terms of the form $\phi |\vec A |$, $|\vec A | |\vec X |$, $\phi  |\vec X |$ can be neglected. Finally, considering a relativistic approximation, i.e, $(\omega^2+{\partial_z}^2)\approx{2\omega(\omega -i\partial_z)}$, we find the following equations\footnote{We work in the gauge $\partial_i A_i=0$ and $A_0=X_0=0$.}:
\begin{equation}
\begin{pmatrix}
\omega -i\partial_z - \frac{m_{\gamma'}^2}{2\omega}
\begin{pmatrix}
 \sin^2\chi & \sin\chi \cos \chi \\
\sin\chi \cos \chi & \cos^2\chi
\end{pmatrix}
\end{pmatrix}
\begin{pmatrix}
A_{\perp} \\
X_{\bot}
\end{pmatrix}
=0,
\label{perpend}
\end{equation}
\begin{equation}
\begin{pmatrix}
\omega -i\partial_z - \frac{1}{2\omega}
\begin{pmatrix}
{m_{\gamma'}^2}\sin^2\chi & {m_{\gamma'}^2} \sin\chi \cos\chi &  {g B\omega\, \tan^2\chi }\\
{m_{\gamma'}^2\sin\chi \cos\chi}&{m_{\gamma'}^2}\cos^2\chi  & {g B\omega}\, \tan\chi  \\
{g B\omega\, \tan^2\chi } & {g B\omega\, \tan\chi } & {m_{\phi}^2}
\end{pmatrix}
\end{pmatrix}
\begin{pmatrix}
A_{\parallel} \\
X_{\parallel}\\
\phi
\end{pmatrix}
=0.
\label{parallel}
\end{equation}
Where $\parallel$ and $\perp$ are the parallel and perpendicular components of the photon field  to the external magnetic field, respectively. Equations~(\ref{perpend}) and (\ref{parallel}) are of the Schroedinger-type, $i \partial_z \Psi\left(z\right)=H\Psi\left(z\right)$, where Eq.~(\ref{perpend}) is the usual one for a model of HP-photon oscillation.
To solve Eq.~(\ref{parallel}) we introduce a rotation matrix that diagonalises the Hamiltonian, i.e, $R^THR={\mbox diag}\left(\omega_1,\omega_2,\omega_3\right)$, where the eigenvalues are given by: $\omega_1=\omega=k$, $\omega_2=\omega-\Omega-\Delta$ and $\omega_3=\omega-\Omega+\Delta$, with:
\begin{equation}
\Omega\equiv \frac{m_{\gamma'}^2+m_{\phi}^2}{4\omega},\,\,\,\,\,\,\,\,\, \Delta\equiv \frac{gB}{2\cos^2\chi} \sqrt{\sin^2\chi+x^2 \cos^4 \chi}, \,\,\,\, \,\,\,\,\, x\equiv \frac{m_{\gamma'}^2-m_{\phi}^2}{2 g B\omega}. \label{freqx}
\end{equation}
The rotation matrix can be conveniently written in terms of two angles, $\theta$ and $\chi$:
\begin{equation}
R=\begin{pmatrix} \cos\chi & \cos \theta \sin \chi & -\sin\theta \sin \chi\\
-\sin\chi & \cos\theta \cos\chi & -\cos \chi \sin \theta \\ 0 & \sin\theta  & \cos\theta
\end{pmatrix}, \,\, \sin\theta= \frac{\sin\chi}{\sqrt{\mathcal F^2+\sin^2\chi}}, \,\,\,\,\,\,\, \mathcal F=\left(x +\frac{2\Delta}{Bg}\right).
\label{matrizR}
\notag
\end{equation}
The states  $X_{\parallel}$ and $\phi$ are sterile to matter currents. Note that the limit $B\rightarrow0$ can be obtained by taking $\theta=\{0,n\pi\}$ when $m_{\gamma'}>m_{\phi}$ and $\theta=(2n+1)\pi/2$ when $m_{\phi}>m_{\gamma'}$. 

The evolution of the interaction states can be obtained from the evolution of mass eigenstates, related by $\Psi(z)=R\Psi'(z)$, prime fields being mass eigenstates. Solving for both amplitudes of the photon after transversing a region of length $L$ we find:
\begin{eqnarray}
A_\parallel(L)&=& e^{-i\omega L}\left(\cos^2\chi+ \sin^2\chi\left(e^{i(\Omega+\Delta) L} \cos^2\theta+e^{i(\Omega-\Delta)L} \sin^2\theta\right)\right) \label{amplitudepar},\\
A_\perp(L)&=&e^{-2iL\omega}\cos^2\chi\sin^2\chi\left(1-e^{iLm_{\gamma'}^2/(2\omega)}\right)^2.
\label{amplitudeperp}
\end{eqnarray}

\section{Ellipticity and rotation effects}
After transversing the magnetic region $L$, the beam has changed its amplitude and phase as we see in Eqs.~(\ref{amplitudepar})-(\ref{amplitudeperp}), meaning that the beam develops a small ellipticity component and a rotation of the polarisation plane. Thus, the amplitudes evolve according to $A_{\parallel,\perp}(z)\propto(1-\epsilon_{\parallel,\perp}(z)) e^{-i\omega z+i\varphi_{\parallel, \perp}(z)}$, the constant of proportionality being the initial polarisation angle, $\alpha_0$, with respect to the direction of the magnetic field $\vec{B}$.  The change in the ellipticity and rotation angles it is given, respectively, by $\psi=\sin(2\alpha_0) \left(\varphi_\parallel -\varphi_\perp\right)/2$ and $\delta \alpha=\sin(2\alpha_0) \left(\epsilon_\parallel- \epsilon_\perp\right)/2$ . After some manipulation of $A_\parallel (z)$ we find:
\begin{eqnarray}
\varphi_\parallel&=&  \sin^2\chi \left[\sin(\Omega z) \cos(\Delta z)+\cos (2\theta) \sin(\Delta z) \cos(\Omega z)\right] \label{parallelellip},\\
\epsilon_\parallel&=& \sin^2\chi \left[1-\cos(\Omega z)\cos(\Delta z)+\cos(2\theta) \sin(\Omega z) \sin(\Delta z)\right]. \label{parallelrot}
\end{eqnarray}
A similar analysis for $A_{\perp}$ gives $\varphi_\perp=  \sin^2\chi \sin\left(m_{\gamma'}^2 z/2\omega\right)$ and $\epsilon_\perp=2 \sin^2\chi \sin^2\left(m_{\gamma'}^2 z/4\omega\right)$.

\begin{figure}[t]
\centerline{
\includegraphics[width=0.5\textwidth]{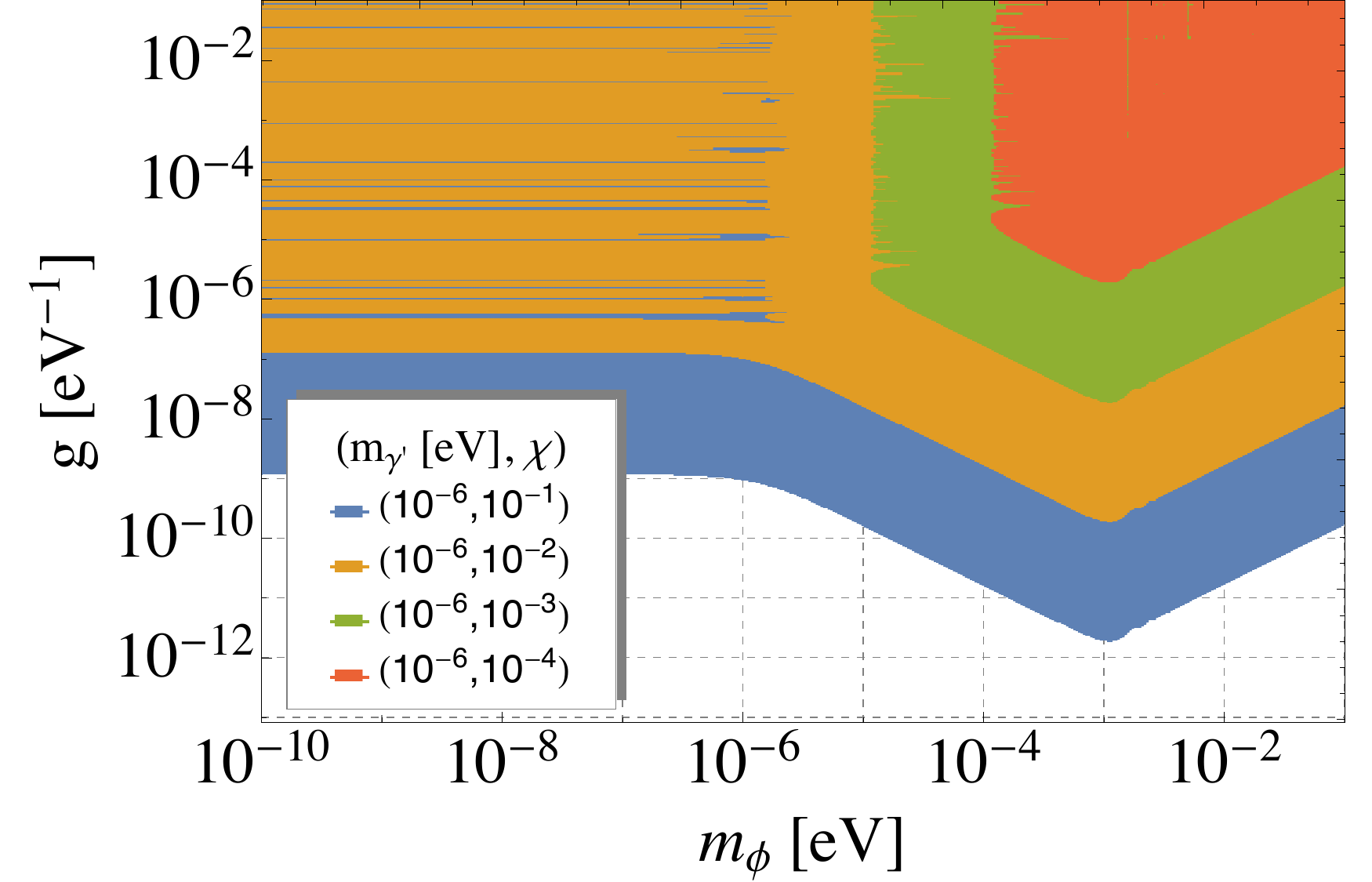}
\includegraphics[width=0.5\textwidth]{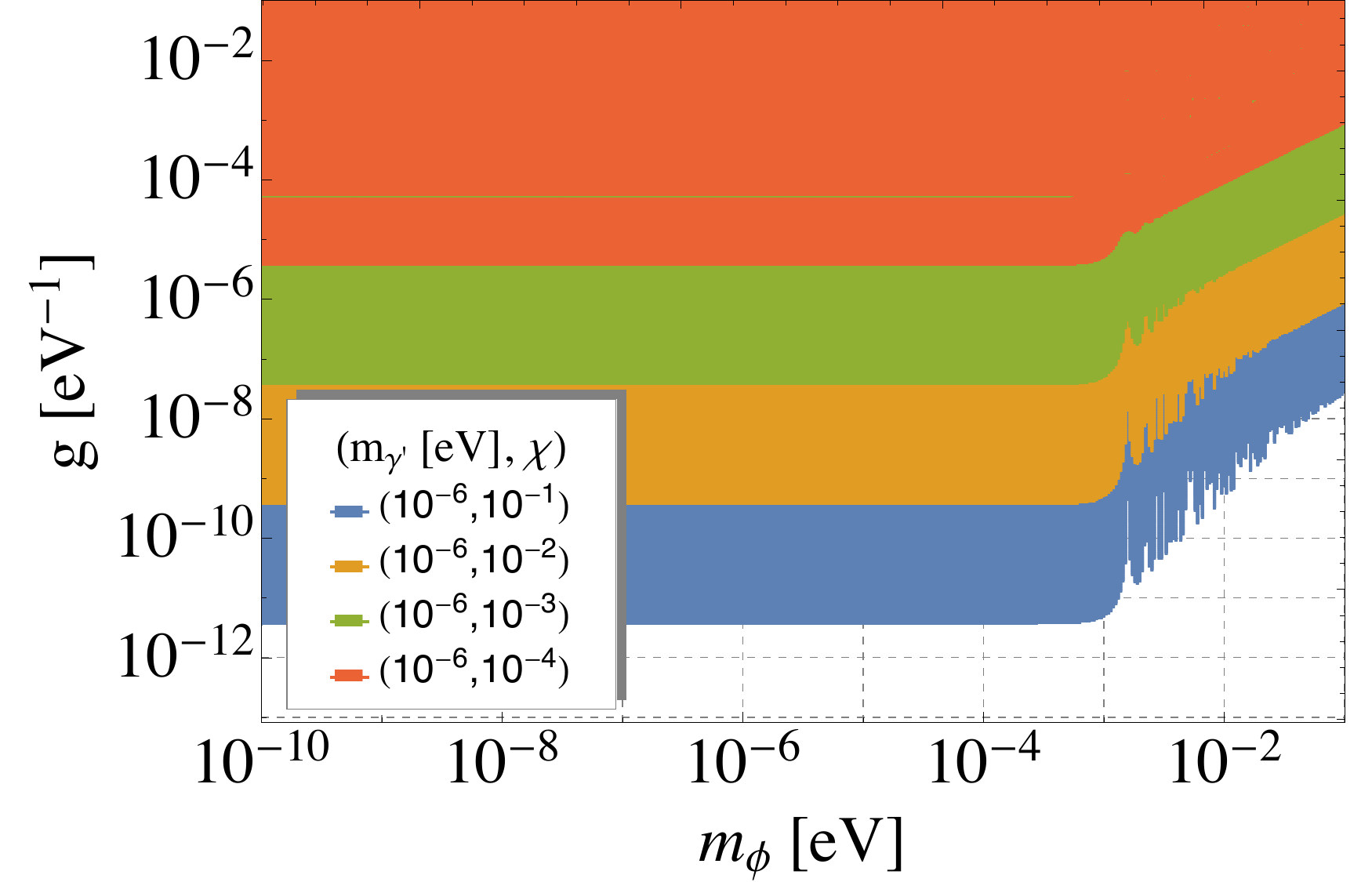}}
\caption{\small{l.h.s. ellipticity constraints on the ALP and  r.h.s. using rotation measurements, for different HP parameters. We have considered $|\delta\alpha_{exp}|=5.2\times10^{-10}$ and $|\psi_{\rm exp}|=9\times10^{-11}$. Both figures assume benchmark values $B=2.5$ T, $L=1$ m, $\omega=1$ eV and $\alpha_0=45^{\circ}$. }}
\label{examples}
\end{figure}

Let us point out that the dimensionless parameter $x$, defined in Eq.~(\ref{freqx}), can be used to define two different regimes: $|x|\ll \chi$, which translates into $\theta\rightarrow \pi/4$ and $|x|\gg \chi$, which translates into $\theta\rightarrow 0$, if $m_{\gamma'}>m_\phi$, or $\theta\rightarrow \pi/2$, if $m_{\gamma'}<m_\phi$. In Fig.~(\ref{examples}) we present exclusion plots to the ALP parameters using both ellipticity and rotation measurements.

{\it Ellipticity effects:}
 Let us first focus on the small mass region, where $|x|\ll1$ and thus, $\theta \sim \pi/4$: this parameter space can remain uncovered if: {\it i)} there is a cancellation between the parallel and perpendicular contributions $\varphi_\parallel-\varphi_\perp \sim 0$, which happens if both $\Delta L$ and $\Omega L \ll 1$. {\it ii)} $\Delta L\gg 1$ and $\Omega L\ll 1$ and $\psi=3\chi^2{m_{\gamma'}^2}L/({8\omega})<|\psi_{exp}|$, where we take $|\psi_{exp}|\sim 9\times 10^{-11}$ as benchmark measured value of the ellipticity angle, as suggested by \cite{pvlas}. For instance, for $\chi=10^{-1}$, the opposite holds, meaning $3\chi^2{m_{\gamma'}^2}L/({8\omega})>|\psi_{ exp}|$, and therefore the small mass region can be constrained up to $g\approx10^{-9}$~eV$^{-1}$, see Fig.~(\ref{examples}). Going below those values of $g$ it is not possible due to the cancellation explained in {\it i)}.  On the other hand, for $\chi=10^{-2}$ the opposite aforementioned condition also holds, but we see  some stripes or gaps in sensitivity in the low mass region. They appear because when the condition $g=4\pi n/(B\chi L)$, where $n\in Z$ is fulfilled, the ellipticity angle drops below $|\psi_{exp}|$.  These gaps in sensitivity can be covered either by changing slightly any of the parameters: $B,L, \omega$. 
Finally, for masses $m_{\gamma'}\gtrsim 10^{-5}$~eV, the condition $|x|\ll \chi$ is no longer fulfilled and the angle $\theta $ starts slowly to approach to $\pi/2$ as $ m_\phi$ grows over $m_{\gamma'}$.  When $\chi \ll |x|$, the expression for the ellipticity angle is well approximated by
\begin{equation}
\psi \propto\chi^2\frac{g^2B^2\omega^2}{m_{\phi}^4}\left(-\frac{m_{\phi}^2 L}{2\omega}+ \sin(\frac{m_{\phi}^2 L}{2\omega})\right),
\label{aproxvar}
\end{equation}
This is almost the same expression of the ellipticity angle for the photon-ALP model (see e.g. \cite{Raffelt:1987im}), but with the replacement of  $g\chi^2$  by the ALP to photon coupling $g_{\phi\gamma\gamma}$. If $m_{\gamma'} > m_\phi$, then the above equation changes, replacing $m_\phi \rightarrow m_{\gamma'}$ and an overall minus sign. Eq.~(\ref{aproxvar}) explains the already familiar {\it V} shape in the mass region $m_\phi \gtrsim 10^{-4}$~eV. 

{\it Rotation effects:}
In the low mass region $m_\phi, m_{\gamma'} \lesssim 10^{-6}$~eV, the conditions $|x|\ll \chi$ and  $\Omega L\ll 1$ hold, thus we can approximate $\epsilon_\parallel\approx 2\chi^2\sin^2\left(g B\chi L/4\right)$ and $\epsilon_\perp\sim \chi^2 m_{\gamma'}^4 L^2/(8\omega)\sim 0$. Therefore, in the low mass region the rotation angle $\delta \alpha$ is mass independent. This gives us the smallest $g$ to be constrained as

\begin{equation}
g\leq \frac{2\sqrt{2|\delta\alpha_{\rm exp}|}}{BL\chi^2}=2.5\times 10^{-12}\,{\rm GeV}^{-1} \left[\frac{2.5\,\rm{T}}{B} \frac{1\,{\rm m}}{L}\left(\frac{|\delta \alpha_{\rm exp}|}{5.2\times 10^{-10}}\right)^{1/2} \left(\frac{10^{-1}}{\chi}\right)^2 \right].
\end{equation}
As the mass grows, we move to the weak mixing regime $\chi \lesssim |x|$, where the ALP starts to decouple from the photon and HP. The change in the polarisation plane can be well approximated as (for $m_{\gamma'}<m_\phi$)
\begin{equation}
\delta \alpha \propto 2\chi^2\left(\sin^2\left(\frac{\Omega L-\Delta L}{2}\right)-\sin^2\left(\frac{m_{\gamma'}^2 L}{2\omega}\right)\right).
\end{equation}
When  $|x|\gg 1$ the difference  $\Omega L-\Delta L$ goes to $m_{\gamma'}^2 L/(2\omega)$ and the right hand side of the above equation cancels. This is the limit where the ALP decouples, and therefore the rotation effect is only due to the HP. 

\section{Conclusions}
In this work we have considered a model that mixes photons, ALPs and HPs, we have shown  interesting  features on observable effects, in this case rotation of the polarisation plane and ellipticity of the beam. The parameters of the model  can still be reasonably constrained using existing results from laboratory experiments. The next step is to consider more stringent scenarios, such as stellar production and early universe. 

{\it Acknowledgments:} This work has been supported by FONDECYT project 1161150.
 

\begin{footnotesize}

\end{footnotesize}


\end{document}